\documentclass{article}

\PassOptionsToPackage{numbers, compress}{natbib}
\usepackage[preprint]{neurips_2026}

\usepackage[utf8]{inputenc} % allow utf-8 input
\usepackage[T1]{fontenc}    % use 8-bit T1 fonts
\usepackage{hyperref}       % hyperlinks
\usepackage{url}            % simple URL typesetting
\usepackage{booktabs}       % professional-quality tables
\usepackage{amsfonts}       % blackboard math symbols
\usepackage{nicefrac}       % compact symbols for 1/2, etc.
\usepackage{microtype}      % microtypography
\usepackage{xcolor}         % colors

\usepackage{comment}

\usepackage{amsmath}
\usepackage{amssymb}
\usepackage{mathtools}
\usepackage{amsthm}
\usepackage{multirow}
\usepackage{graphicx}
\usepackage{colortbl}
\usepackage{subcaption}
\usepackage[numbers]{natbib}
\usepackage{wrapfig}

\title{Music Transcription with (Almost) No Supervision}

\author{%
  \textbf{Saebyeol Shin}\thanks{\texttt{Correspondence to <ss4333@cornell.edu>}} \quad
  \textbf{Chao Wan} \quad
  \textbf{Zhenzhen Liu} \quad
  \textbf{Justin Lovelace} \\[4pt]
  \textbf{Daniel C. Lin} \quad
  \textbf{Kilian Q. Weinberger} \quad
  \textbf{John Thickstun} \\[9pt]
  Cornell University, Ithaca, NY
}

\begin{document}

\maketitle

\begin{abstract}
Competitive music transcription models require large amounts of paired audio-score data, which is scarce due to collection costs, alignment difficulty, and copyright restrictions. Meanwhile, vast quantities of unpaired audio recordings and symbolic scores are freely available but have gone unused. We adopt a cycle-consistent translation framework in which a small amount of paired data acts as a minimal anchor, unlocking the full potential of the unpaired pool. We find that: unpaired data yields surprisingly large gains, especially under limited supervision; unpaired audio contributes more than unpaired scores; incorporating unlabeled audio from a new instrument during training improves transcription for that instrument without any paired supervision. Together, these results suggest that scaling unpaired data offers a practical path toward high-quality transcription for instruments where labeled data remains scarce. \footnote{Code is available at: \url{https://github.com/SaebyeolShin/almost_unsupervised_amt}}
\end{abstract}

\section{Introduction}
Supervised deep learning has achieved impressive results for piano transcription, producing models that approach human-level performance \cite{hawthorne2017onsets,kong2020high,kelz2019deep}. These models learn to translate audio spectrograms to symbolic scores by training on large datasets of aligned audio-score pairs. For piano music, datasets like MAESTRO \cite{hawthorne2018enabling} have been constructed using a computer-monitored Disklavier piano, which is wired to synchronously record a transcript of key presses and releases during an acoustic musical performance, yielding hundreds of hours of perfectly aligned audio-score pairs for supervised learning. However, most instruments and ensembles lack such recording infrastructure, making aligned data scarce. Manual alignment is prohibitively labor-intensive: annotators must identify note onsets and offsets in audio spectrograms, and match them to notes in a score, requiring both musical expertise and painstaking, frame-level precision. Copyright compounds the problem: legal restrictions largely prevent the distribution of aligned audio as research datasets. Meanwhile, unpaired audio and symbolic scores have been accumulating for well over a century: recordings exist since the invention of the gramophone in 1887, and symbolic scores for far longer. Yet harnessing this abundance of data for transcription has remained an open problem. This raises a fundamental question: \textbf{can the vast stores of unpaired audio and symbolic scores — freely available but so far unusable for transcription — be turned into an effective training signal?}

Learning from unpaired cross-modal data resembles the challenge faced by early scholars deciphering ancient scripts: without a Rosetta Stone providing direct translation pairs, alignment must be inferred from structural correspondences and consistency constraints. Cycle consistency \cite{zhu2017unpaired} has proven effective as a self-supervisory signal for learning mappings between unpaired domains, but audio-to-symbolic transcription involves a fundamental modality shift, from continuous spectrograms to discrete event sequences, that makes direct application non-trivial. 

To address this challenge, we learn a cycle-consistent mapping between CQT spectrograms and a continuous latent space derived from a pre-trained score VAE. This latent representation provides a smooth intermediate space for cross-modal translation while preserving score structure. However, cycle consistency alone is insufficient: without paired supervision, the learning objective admits degenerate but cycle-consistent solutions, such as globally pitch-shifted transcriptions. We find that a small amount of paired data (as little as 1.6 hours) acts as a minimal anchor that resolves this alignment ambiguity and enables the much larger pool of unpaired data to contribute meaningfully to learning. 
Our primary focus is on what happens after this anchor is in place: how unpaired data scales performance across supervision regimes, which modality contributes more strongly to learning, and whether unlabeled target-domain audio alone suffices for domain adaptation.

\begin{figure*}[t]
    \centering
\includegraphics[width=\linewidth]{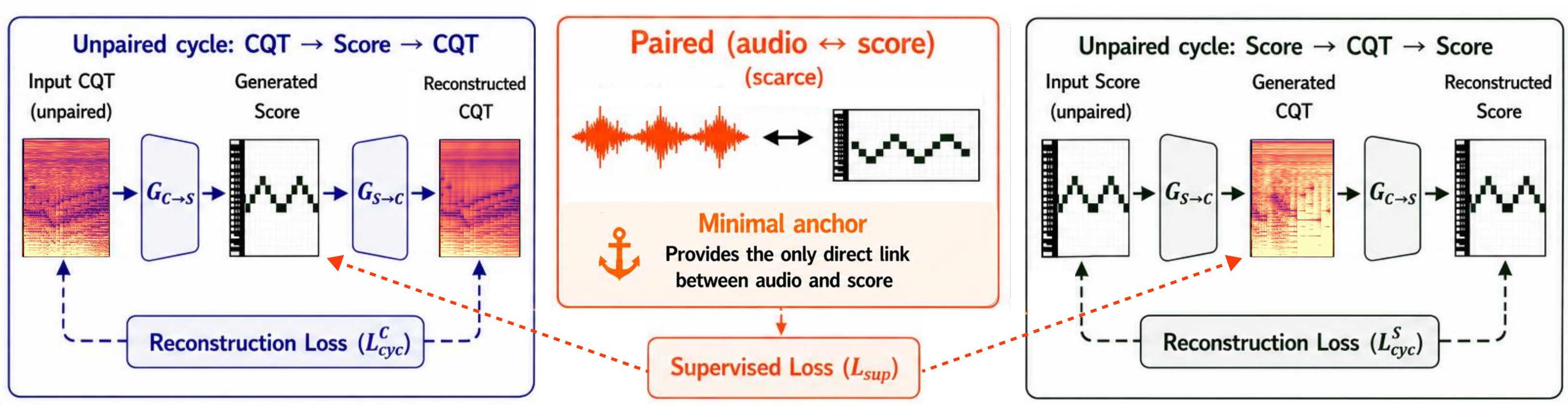}
    \caption{
    We show the value of unpaired data for music transcription under limited paired supervision. While paired audio–score data is scarce, large amounts of unpaired audio and score data are available. We find that a small amount of paired data acts as a minimal anchor that enables unpaired data to contribute effectively, leading to substantial gains in low-resource regimes.
}
    \label{fig:firstfig}
    \vspace{-8pt}
\end{figure*}

We find that unpaired data provides consistent and substantial gains on both MAESTRO~\cite{hawthorne2018enabling} and GuitarSet~\cite{xi2018guitarset}, with the largest improvements in low-resource regimes where paired supervision is scarce. In controlled experiments on MAESTRO, training a model on 1.6h of paired data achieves 66.93 Frame F1; adding $100×\times$ more unpaired data raises this to 75.45, compared to 87.43 for a model trained on 161h of fully supervised data.
We further find that unpaired audio provides stronger learning signal than scores at matched budgets, suggesting that acoustic diversity is the primary driver of learning from unpaired data. 
Finally, incorporating unlabeled audio from an unseen instrument during training improves transcription of that instrument without any paired target-domain supervision: our model achieves 64.81 Frame F1 on GuitarSet \cite{xi2018guitarset} using only unlabeled guitar audio alongside minimal piano supervision, exceeding a fully supervised out-of-domain baseline. 
We also validate these findings in a multi-instrument setting on MusicNet-EM \cite{thickstun2016learning}, where unpaired data from mismatched sources, held-out scores and out-of-domain audio consistently improves transcription under minimal paired supervision, demonstrating that the benefits of cycle-consistent training extend beyond single-instrument settings. These findings suggest that the long-standing abundance of unpaired audio offers a practical and largely untapped path toward high-quality transcription for instruments where labeled data remains scarce.

\section{Related Work}

\subsection{Automatic Music Transcription}
Previous work on music transcription methods can be classified in two distinct representational paradigms. Much of the early work on transcription predicted rasterized, time-frequency matrix representations of scores with a structural similarity to images \cite{poliner2006discriminative,boulanger2012modeling,kelz2016potential}. The Onsets \& Frames model~\cite{hawthorne2017onsets} advanced this score-based approach by jointly predicting onset events and frame-wise activations through separate neural stacks, achieving strong transcription performance. Subsequent refinements improved temporal modeling \cite{kelz2019deep} and added pedal prediction \cite{kong2020high}. Despite architectural advances, these methods share a common characteristic: they predict dense, two-dimensional representations of scores. More recently, Hawthorne, et al. \cite{hawthorne2021sequence} suggested that transcription be considered as a sequence-to-sequence problem, employing Transformer that generates scores in a discrete, language-inspired format. MT3 \cite{gardner2021mt3} further generalized this paradigm to multi-instrument transcription, demonstrating that a single Transformer model can jointly transcribe diverse instrument combinations across datasets and improve cross-dataset generalization. These methods use autoregressive token generation instead of dense score prediction, which allows for flexible structured outputs but also adds sequential sampling to the decoding process.

Our work intentionally returns to the frame-wise score formulation. This raster matrix representation encodes pitch activity as a dense two-dimensional array over time, making it naturally suited for convolutional autoencoders. The resulting latent space is smooth and spatially structured, enabling gradient-based cycle consistent learning. In contrast, autoregressive token sequences lack this spatial regularity and would require additional design choices to produce an equivalently structured latent representation. 
In contrast to previous work, which fundamentally relies on supervised learning using paired audio-score data, our work learns to transcribe audio using predominantly unpaired data. While MT3 demonstrates transfer across datasets, it requires large paired datasets for initial training, and additional paired data for transfer. Unlike MT3 and related methods, our approach requires only 1.6 hours of paired supervision as a minimal anchor, and shows that the remaining data budget is better spent on unpaired audio, including unlabeled audio from unseen instruments, than on additional paired labels.

\subsection{Unpaired and Semi-Supervised Learning}
\textbf{Low-supervision and unsupervised learning in music transcription.}
Recent work has addressed low-resource and out-of-distribution settings in AMT through semi-supervised learning and domain adaptation. ReconVAT \cite{cheuk2021reconvat} uses unlabeled audio as additional training signal through consistency regularization and reconstruction losses to improve generalization, while Riley et al. \cite{riley2024high} transfer piano models to guitar using score-activation alignment. Sato et al. \cite{sato2024annotation} combine synthetic pre-training with adversarial domain confusion to adapt to real audio without labels, and Edwards et al. \cite{edwards2024data} demonstrate that large-scale augmentation improves robustness to distribution shift. Liu et al. \cite{liu2025unsupervised} propose teacher-student learning with cross-version consistency for unsupervised domain adaptation across instruments. 
Earlier work has also explored transcription with little or no paired supervision. Berg-Kirkpatrick et al. \cite{berg2014unsupervised} proposed a probabilistic framework for unsupervised piano transcription, while Choi and Cho \cite{choi2019deep} introduced an unsupervised drum transcription system trained through differentiable audio reconstruction. However, these approaches either focus on highly structured settings or rely on explicit generative priors within the audio domain, rather than learning cross-modal translation between unpaired audio and symbolic score representations. 
While prior methods reduce reliance on labeled target-domain data, they still fundamentally depend on synthetic labels, pseudo-labels, score alignment, or domain-specific reconstruction priors. Our work instead investigates whether transcription mappings can be learned directly from predominantly unpaired audio and symbolic data, without synthetic rendering, pseudo-labeling, or score alignment. Crucially, none of these methods unlock the vast stores of completely unpaired audio and symbolic data, that form the bulk of what is freely available.
\\\\
\noindent\textbf{Unsupervised Domain Translation.} 
Unsupervised domain translation aims to learn mappings between two domains without paired data. Most progress in this area has been made in the image modality, where adversarial translation \cite{zhu2017unpaired, kim2017learning, yi2017dualgan, liu2017unsupervised} is a representative approach, introducing cycle consistency as a constraint for learning bidirectional mappings between unpaired domains, showing effectiveness in tasks like style transfer and domain adaptation. In contrast, applications in audio and music remain limited. In speech, prior work mostly focuses on voice conversion and speech style transfer \cite{kaneko2018cyclegan, qian2019autovc}. In music, cycle-consistent objectives have been applied to musical audio transformation in time–frequency representations. For example, TimbreTron \cite{huang2018timbretron} applies cycle-consistent training to perform musical timbre transfer in the log-CQT domain. Similar ideas have also been explored for symbolic music domain transfer, such as for translating genres and styles in the piano-roll/score domain \cite{brunner2018symbolic}. To the best of our knowledge, we are the first work to study the use of such techniques for cross-modal translation in the music domain.

\begin{figure*}[t]
    \centering
    \includegraphics[width=1\textwidth]{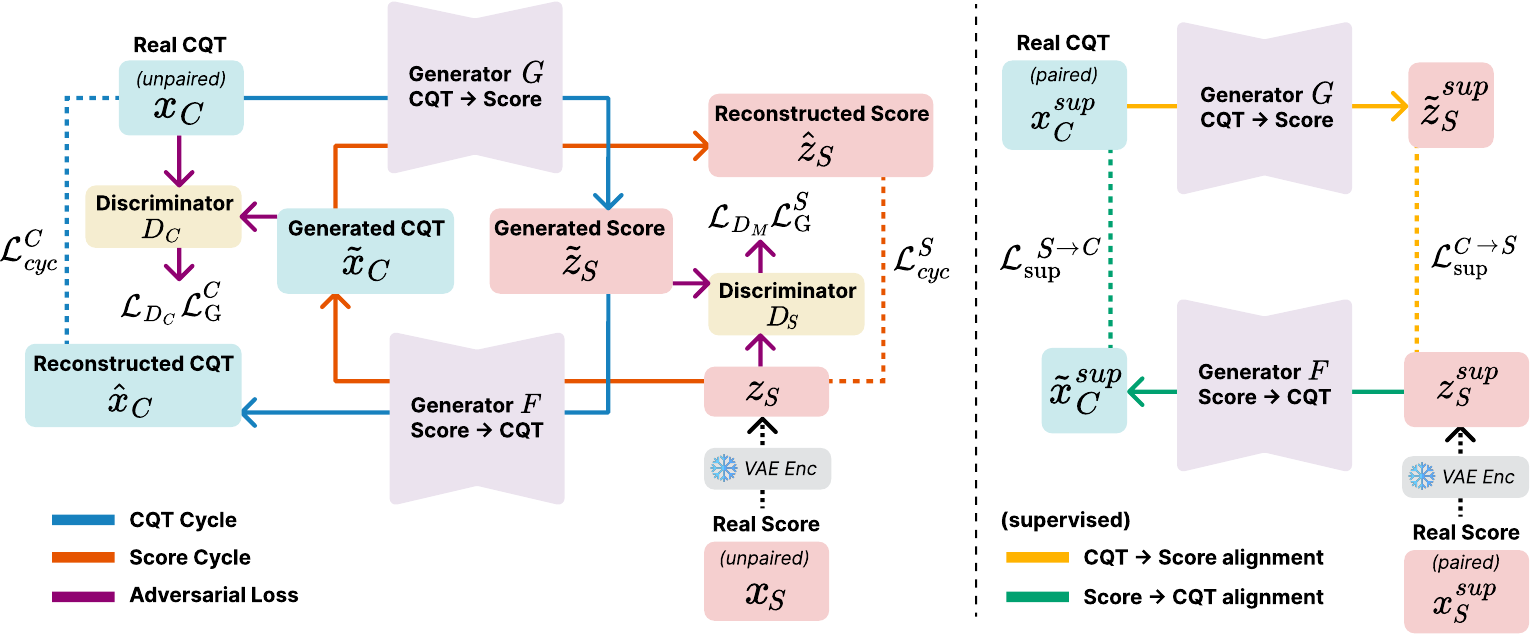}
    \caption{
    Overview of the proposed semi-supervised latent-space transcription framework. Two generators: a CQT-to-score transcriber and a score-to-CQT synthesizer are shared across paired and unpaired training paths. A small amount of paired audio-score data anchors the cross-modal mapping, while abundant unpaired data contributes additional learning signal via cycle consistency. This combination prevents degenerate pitch-shift solutions and enables transcription with predominantly unpaired data.}
    \label{fig:main}
\end{figure*}
\section{Cycle Consistent Transcription}

We adopt a cycle-consistent translation framework to make the unpaired pool usable for music transcription, turning audio and scores that cannot be directly supervised into an effective training signal. We treat the \textit{paired:unpaired} ratio as a controlled variable, holding the total available data fixed and varying only how much of it is used as paired supervision.

We consider mappings between CQT spectrograms $X_C$ and symbolic scores $x_S$, with data distributions $X_C \sim p_C(x_C)$ and $x_S \sim p_S(x_S)$. To facilitate stable training, we introduce a continuous latent space $\mathcal{Z}_S$ obtained from a pre-trained score VAE with encoder $E_S$, writing $z_S = E_S(x_S)$, and learn mappings (Figure~\ref{fig:main}):
\[
G: X_C \to \mathcal{Z}_S,
\qquad
F: \mathcal{Z}_S \to X_C.
\]

\noindent\textbf{Pre-trained Variational Autoencoder.}
We map symbolic scores into a continuous latent space $\mathcal{Z}_S$ using a pre-trained VAE. 
This latent representation provides a smooth and differentiable space for cross-modal translation, bypassing the need to backpropagate through discrete symbolic representations directly.
The VAE is trained with the standard ELBO objective $\mathbb{E}[-\log p(x_S \mid z_S)] + \beta D_{\mathrm{KL}}(q(z_S \mid x_S)|p(z_S))$. After pre-training, the encoder $E_S$ is frozen and used to provide stable latent representations for adversarial training.

\noindent\textbf{Adversarial and Cycle Consistency Objectives.}
We adopt standard cycle-consistency objectives~\cite{zhu2017unpaired} with LSGAN losses~\cite{mao2017least}. Discriminators $D_S$ and $D_C$ distinguish real from generated samples in each domain. Cycle consistency enforces that composing forward and backward mappings approximates the identity:
\begin{align}
\mathcal{L}_{\text{cyc}}^C &=
\mathbb{E}_{x_C}\bigl[\lVert F(G(x_C)) - x_C \rVert_1\bigr], \\
\mathcal{L}_{\text{cyc}}^S &=
\mathbb{E}_{x_S}\bigl[\lVert G(F(z_S)) - z_S \rVert_1\bigr].
\end{align}

\noindent\textbf{Cycle-Aware Feature Matching.}
Standard feature matching~\cite{kumar2019melgan, wang2018high} compares discriminator features between real and generated samples across domains. In cross-modal translation, this is problematic: real and translated samples \emph{should} differ, so minimizing their feature distance suppresses the very transformation we want to learn. We instead apply feature matching within each domain, between real samples and their cycle-reconstructions:
\begin{align}
\mathcal{L}_{\text{FM}}^C &= \sum_{i=1}^{L} \frac{1}{N_i}
\bigl\lVert D_C^{(i)}(x_C) - D_C^{(i)}(F(G(x_C))) \bigr\rVert_1,
\end{align}
where $L$ is the number of features. The symmetric loss $\mathcal{L}_{\text{FM}}^S$ is defined analogously. 

\noindent\textbf{Anchoring with Paired Supervision.}
Cycle consistency enforces approximate invertibility of the learned mappings but does not uniquely determine alignment between domains. In transcription, this leads to a non-identifiability: multiple mappings, including globally pitch-shifted solutions, satisfy the training objective. As a result, models trained with unpaired data alone may converge to pitch-misaligned solutions. 
A small amount of paired supervision resolves this ambiguity, providing the anchor signal that allows unpaired data to contribute meaningfully. At each training step, we sample one batch from the unpaired pool for adversarial, cycle consistency, and feature matching losses, and one batch from a fixed paired subset for direct supervised losses:
\begin{align}
\mathcal{L}_{\text{sup}}^{C \to S} &=
\mathbb{E}_{(x_C, x_S)}\bigl[\lVert G(x_C) - z_S \rVert_1\bigr], \\
\mathcal{L}_{\text{sup}}^{S \to C} &=
\mathbb{E}_{(x_C, x_S)}\bigl[\lVert F(z_S) - x_C \rVert_1\bigr].
\end{align}
The reverse loss $\mathcal{L}_{\text{sup}}^{S \to C}$ prevents $F$ from drifting independently of $G$ during early training. The proportion of paired data is a controlled variable in our experiments; we study how the gain from unpaired data varies as this anchor grows.

\noindent\textbf{Full Training Objective.}
Generators minimize:
\begin{align}
\mathcal{L}_G =
\mathcal{L}_G^S + \mathcal{L}_G^C
+ \lambda_{\text{fm}}(\mathcal{L}_{\text{FM}}^C + \mathcal{L}_{\text{FM}}^S)
+ \lambda_{\text{cyc}}(\mathcal{L}_{\text{cyc}}^C + \mathcal{L}_{\text{cyc}}^S)
+ \lambda_{\text{sup}}(\mathcal{L}_{\text{sup}}^{C \to S} +
  \mathcal{L}_{\text{sup}}^{S \to C}).
\end{align}
Discriminators minimize $\mathcal{L}_D = \mathcal{L}_{D_S} + \mathcal{L}_{D_C}$ with alternating updates. We use $\lambda_{\text{cyc}}=5.0$, $\lambda_{\text{fm}}=1.0$, $\lambda_{\text{sup}}=1.0$, image pools of size 128, and sample 50\% of discriminator examples from the pool.
\section{Experimental Methodology}
\begin{table*}[tb]
\centering
\caption{Transcription performance on MAESTRO. For each unpaired budget, \textit{paired-only} and \textit{paired + unpaired} training are compared 
to isolate the contribution of unpaired data. The unpaired-only results reflect the instability of purely unsupervised training: some runs converge 
to reasonable solutions while others have pitch shift. The \textit{paired:unpaired} ratio (e.g., 1\,:\,100 = 100$\times$ more unpaired than paired) is shown alongside absolute durations (train split total = 161.1 h).}
\begin{minipage}[c]{0.6\textwidth}
\centering
\resizebox{\linewidth}{!}{
\begin{tabular}{@{}clc@{}}
\toprule
\textbf{Paired : Unpaired} & \textbf{Training Setup} & \textbf{Frame F1} \\
\midrule
\multirow{2}{*}{\begin{tabular}[c]{@{}c@{}}0 : 1\\ \footnotesize\color{gray}{(0 h : 161.1 h)}\end{tabular}}
  & Unpaired only (shifted)  & 13.77 \\
  & Unpaired only (stable)   & 68.90 \\
\midrule
\multirow{2}{*}{\begin{tabular}[c]{@{}c@{}}1 : 1000\\ \footnotesize\color{gray}{(9.7 min : 160.9 h)}\end{tabular}}
  & Paired-only        & 47.24 \\
  & Paired + Unpaired  & 71.37 \\
\midrule
\multirow{2}{*}{\begin{tabular}[c]{@{}c@{}}1 : 100\\ \footnotesize\color{gray}{(1.6 h : 159.5 h)}\end{tabular}}
  & Paired-only        & 66.93 \\
  & Paired + Unpaired  & 75.45 \\
\midrule
\multirow{2}{*}{\begin{tabular}[c]{@{}c@{}}1 : 19\\ \footnotesize\color{gray}{(8.1 h : 153.0 h)}\end{tabular}}
  & Paired-only        & 75.62 \\
  & Paired + Unpaired  & 79.34 \\
\midrule
\multirow{2}{*}{\begin{tabular}[c]{@{}c@{}}1 : 9\\ \footnotesize\color{gray}{(16.1 h : 145.0 h)}\end{tabular}}
  & Paired-only        & 79.58 \\
  & Paired + Unpaired  & 81.81 \\
\midrule
\color{gray}{1 : 0} & \color{gray}{Paired-only} & \color{gray}{87.43} \\
\bottomrule
\end{tabular}
}
\end{minipage}
\hfill
\begin{minipage}[c]{0.38\textwidth}
\centering
\includegraphics[width=\linewidth]{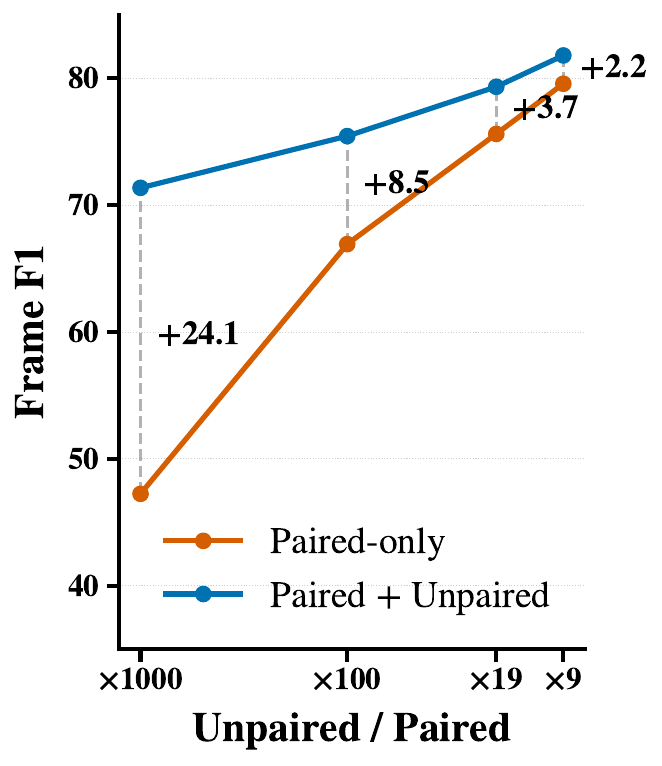}
\end{minipage}
\label{tab:main_results}
\vspace{-5pt}
\end{table*}

We conduct experiments to evaluate our approach across varying amounts of paired supervision. We assess transcription quality on piano using MAESTRO, analyze how performance scales with paired data availability, evaluate cross-instrument generalization using GuitarSet, and evaluate multi-instrument transcription using MusicNet-EM.

\textbf{Datasets.} \textbf{MAESTRO}~\cite{hawthorne2018enabling} contains $\approx$200 hours of aligned piano audio and scores; we sample $\{0.1, 1, 5, 10\}\%$ of the training set as paired supervision and treat the remainder as unpaired. This corresponds to paired-to-unpaired ratios of $\{1\!:\!1000, 1\!:\!100, 1\!:\!19, 1\!:\!9\}$. \textbf{GuitarSet}~\cite{xi2018guitarset} contains 360 guitar recordings used for zero-shot cross-instrument evaluation and as unlabeled target-domain audio in domain adaptation experiments. \textbf{MusicNet-EM}~\cite{thickstun2016learning, maman2022unaligned} is a multi-instrument classical music dataset; we use 3 recordings (0.47\,h) as paired supervision, chosen to cover all 11 instrument classes in the dataset, 305 recordings as unpaired scores, and 532 publicly available recordings ($\approx$303\,h) from the \textbf{Isabella Stewart Gardner Museum} as unpaired audio. Full preprocessing and split details are provided in Appendix~\ref{app:data}.

\textbf{Baselines.} We compare three training configurations: 
\textit{paired-only} (supervised losses only, no unpaired data), 
\textit{unpaired-only} (cycle consistency alone, no paired data), and 
\textit{paired + unpaired} (our semi-supervised model with different \textit{paired:unpaired} ratios). The 100\% paired-only model serves as the fully supervised upper bound. Full training details are in Appendix~\ref{app:training}.

\textbf{Metrics.} We report two metrics. \textit{Frame F1} measures frame-level pitch accuracy at 50\,frames per second (20\,ms resolution) over binary pitch vectors of size 88. \textit{Multi-Instrument Frame F1} additionally requires correct instrument channel assignment.

\begin{figure*}[t]
\centering
\includegraphics[width=1\textwidth]{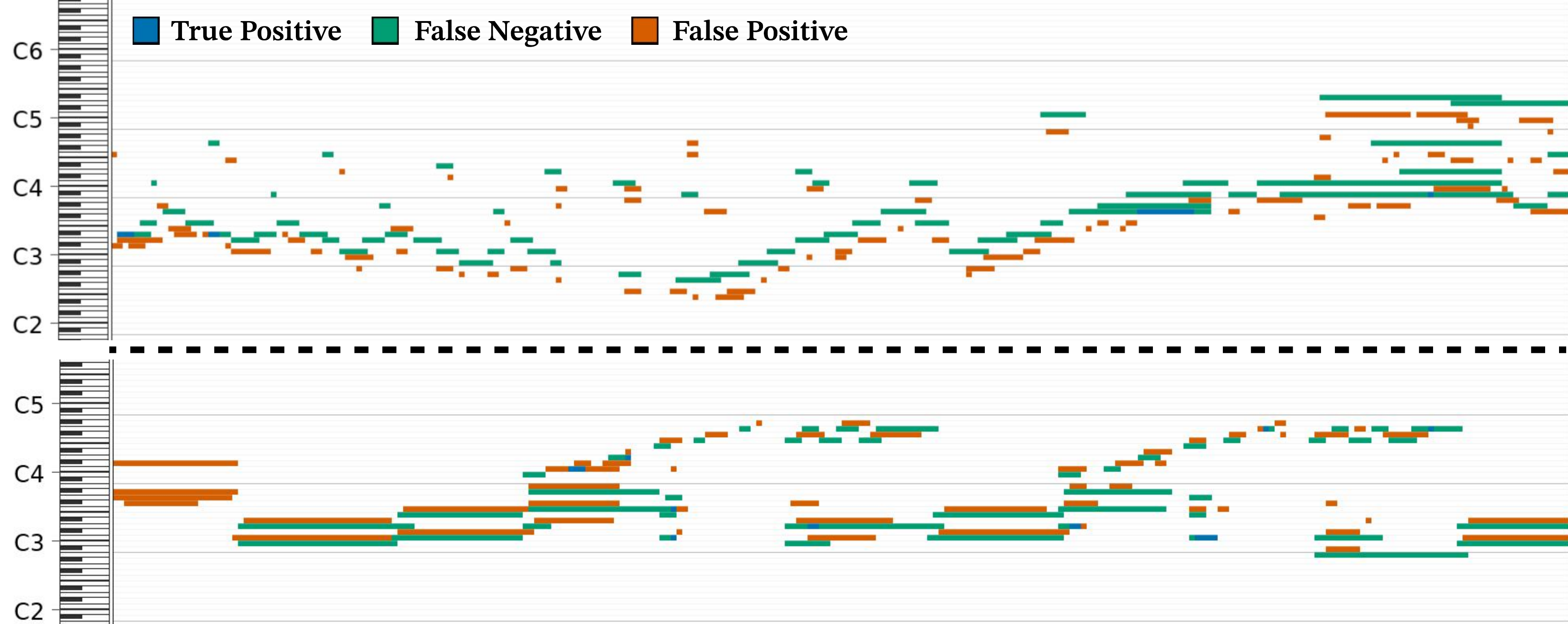}
\caption{Pitch-shift failure modes from two independent unpaired-only training 
runs, where each run converges to a different pitch shift. Despite this failure, the cycle loss remains low since $\text{CQT} \to \text{Score} 
\to \text{CQT}$ reconstruction is unaffected by a globally shifted score.}
\label{fig:collapse}
\vspace{-5pt}
\end{figure*}

\section{Results}
\label{sec:results}
\subsection{Main Results: Scaling with Paired Data}

Table~\ref{tab:main_results} shows transcription performance across varying amounts of paired supervision. The key comparison is between \textit{paired-only} and \textit{paired + unpaired} at each supervision level, isolating the contribution of unpaired data under a fixed paired budget.
\vspace{-10pt}
\\\\
\begin{wrapfigure}{r}{0.5\linewidth}
\vspace{-15pt}
\centering
\includegraphics[width=0.9\linewidth]{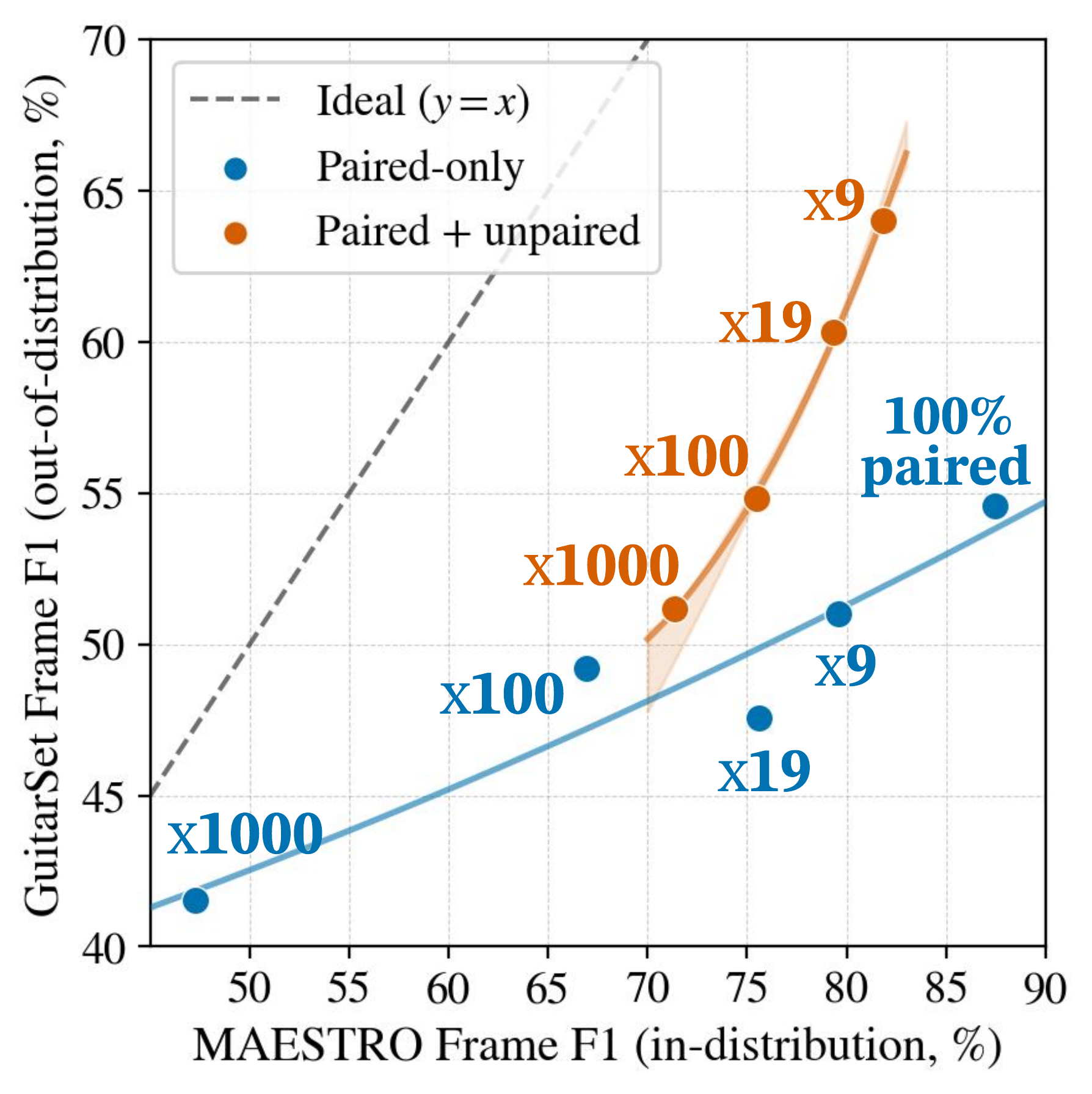}
\caption{Paired + unpaired training is more robust to distribution shift than paired-only training. An ideal robust model (dashed line) performs equally well on the training distribution (MAESTRO) and on a held-out instrument (GuitarSet). Paired + unpaired models consistently shrink this robustness gap across all regimes tested (1\,:\,9 to 1\,:\,1000 paired-to-unpaired).}
\label{fig:idood}
\vspace{-13pt}
\end{wrapfigure}
\noindent\textbf{Unpaired data provides the largest gains when paired supervision is scarce.} When paired supervision is relatively abundant (e.g., \textit{paired:unpaired} ratios of 1:9 and 1:19), adding unpaired data yields only modest improvements, suggesting that paired supervision already provides sufficient anchoring signal. In contrast, when paired data becomes limited, incorporating large amounts of unpaired data substantially improves performance. At a 1:100 ratio (1\% paired), our method achieves 75.45 Frame F1 vs.\ 66.93 for the paired-only baseline (+8.52), recovering 86.3\% of fully supervised performance (87.43 Frame F1). In the most extreme setting, with almost $1000\times$ unpaired data available, adding unpaired data improves Frame F1 from 47.24 to 71.37 (+24.13). These results suggest that once sufficient anchoring is established, additional paired data and cycle consistency provide increasingly overlapping signal, while unpaired data becomes most valuable in extremely low-supervision regimes.
\\\\
\noindent\textbf{Unpaired-only training is unstable.} Without any paired supervision, training exhibits high variance across runs. Some runs converge to reasonable solutions (68.90 Frame F1), while others collapse entirely (13.77 Frame F1) due to a systematic pitch shift: the model learns to predict notes at incorrect pitches, with the direction and magnitude varying unpredictably across runs. Crucially, this failure is difficult to detect by the training objective, since $\text{CQT} \to \text{Score} \to \text{CQT}$ can reconstruct the original spectrogram even with a pitch-shifted score; the cycle loss continues to decrease while transcription F1 deteriorates. We visualize examples of such failure runs in Figure~\ref{fig:collapse}. With only 1.6h of paired supervision, this failure mode is eliminated entirely, enabling consistent convergence across runs.

\begin{table*}[tb]
\begin{minipage}[t]{0.49\linewidth}
\centering
\caption{Effect of unpaired data modality under \textit{paired:unpaired} supervision on MAESTRO. At matched unpaired budgets (1\,:\,90), audio-only outperforms score-only, suggesting that acoustic diversity provides a stronger cycle consistency signal than symbolic coverage alone. Combining both modalities at 1\,:\,100 further improves performance.}
\vspace{-3pt}
\resizebox{1\linewidth}{!}{
\begin{tabular}{@{}lcc@{}}
\toprule
\textbf{Training Setup} & \textbf{Paired : Unpaired} & \textbf{Frame F1} \\ 
\midrule
Paired-only               & 1 : 0  & 66.93 \\
\midrule
+ unpaired audio only     & 1 : 90 & 72.46 \\
+ unpaired score only     & 1 : 90 & 70.51 \\
+ unpaired audio \& score & 1 : 90 & 72.16 \\
\midrule
+ unpaired audio \& score & 1 : 100 & 75.45 \\
\bottomrule
\end{tabular}
}
\label{tab:unpaired_modality}
\end{minipage}
\hfill
\begin{minipage}[t]{0.48\linewidth}
\centering
\caption{Effect of adding unlabeled GuitarSet audio to the unpaired pool, on top of 1\,:\,100 \textit{paired:unpaired} MAESTRO supervision. Without any paired guitar labels, incorporating unlabeled guitar audio substantially improves GuitarSet 
transcription while leaving in-domain piano performance intact.}
\resizebox{0.75\linewidth}{!}{
\setlength{\tabcolsep}{6pt}
\renewcommand{\arraystretch}{1.15}
\begin{tabular}{@{}lc@{}}
\toprule
\textbf{Unpaired Training Data} & \textbf{Frame F1} \\
\midrule
\rowcolor{gray!15}
\multicolumn{2}{c}{\textbf{MAESTRO Test set}} \\
MAESTRO only      & 75.45 \\
+ GuitarSet audio           & 76.23 \\
\rowcolor{gray!15}
\multicolumn{2}{c}{\textbf{GuitarSet Test set}} \\
MAESTRO only      & 54.81 \\
+ GuitarSet audio           & 64.81 \\
\bottomrule
\end{tabular}}
\label{tab:guitarset_train}
\end{minipage}
\end{table*}

\begin{table*}[tb]
\begin{minipage}[t]{0.48\linewidth}
\centering
\caption{Cross-instrument evaluation (MAESTRO $\rightarrow$ GuitarSet). Models are trained on piano and evaluated on guitar without fine-tuning. Paired + unpaired training improves cross-instrument generalization over the paired-only baseline.}
\resizebox{\linewidth}{!}{
\begin{tabular}{@{}lcc@{}}
\toprule
\textbf{Training Setup} & \textbf{Paired : Unpaired} & \textbf{Frame F1} \\ 
\midrule
Paired-only              & 1 : 0   \footnotesize\color{gray}{(1.6 h : 0 h)}    & 49.20 \\
Paired + Unpaired        & 1 : 100 \footnotesize\color{gray}{(1.6 h : 159.5 h)} & 54.81 \\
\midrule
\color{gray}{Paired-only (full)} & \color{gray}{1 : 0 \footnotesize{(161.1 h : 0 h)}} & \color{gray}{54.57} \\
\bottomrule
\end{tabular}
}
\label{tab:ood} 
\end{minipage}
\hfill
\begin{minipage}[t]{0.48\linewidth}
\centering
\caption{Regularization effect of unpaired data. With only 6 minutes of paired data, the model severely overfits (high train F1, low val/test F1). Adding unpaired data reduces overfitting and improves generalization despite not providing explicit alignment signal.}
\resizebox{\linewidth}{!}{
\begin{tabular}{@{}lccc@{}}
\toprule
\textbf{Training Data} & \textbf{Train F1} & \textbf{Val F1} & \textbf{Test F1} \\ \midrule
 6 min paired only& 98.39 & 23.84 & 27.82 \\
+ 60 min unpaired & 86.86 & 42.80 & 43.36 \\ 
\bottomrule
\end{tabular}}
\label{tab:overfit}
\end{minipage}
\end{table*}

\subsection{Domain Adaptation with Unlabeled Target-Domain Audio}
\label{sec:guitarset_train}

The value of unpaired data is not limited to in-domain performance. We show that incorporating unlabeled audio from a new instrument yields substantial gains in transcription of that instrument, outperforming a model trained with full supervision on out-of-domain data.

\noindent\textbf{Unpaired training improves cross-instrument generalization even without target-domain data.} Table~\ref{tab:ood} evaluates models trained exclusively on piano (MAESTRO) directly on GuitarSet without any fine-tuning. Our 1:100 paired:unpaired model achieves 54.81 Frame F1, compared to 49.20 Frame F1 for the paired-only baseline (+5.61), isolating the contribution of cycle-consistent training to cross-instrument robustness. Moreover, it also matches the model trained with all paired data (54.81 vs 54.57 Frame F1). Figure~\ref{fig:idood} visualizes this trend across all data regimes: paired + unpaired models consistently lie closer to the ideal robust line (y\,=\,x) than their paired-only counterparts. We hypothesize that
cycle-consistent training encourages the model to learn pitch representations less entangled with piano-specific timbral cues. 
 \\\\
\noindent\textbf{Incorporating unlabeled target-domain audio outperforms fully supervised cross-domain transfer.} 
These zero-shot gains motivate a stronger intervention: adding unlabeled GuitarSet audio directly to the unpaired pool. Table~\ref{tab:guitarset_train} reports results on both MAESTRO and GuitarSet test sets. All models use 1.6h paired set of MAESTRO as the anchoring signal; only the unpaired audio pool differs. Adding unlabeled GuitarSet audio to the unpaired pool improves guitar transcription from 54.81 to 64.81 Frame F1 (+10.00), exceeding the performance of all models trained only on MAESTRO data, despite no paired guitar labels being available at any point during training. Critically, only one modality from the target domain is required: unlabeled audio alone suffices to bridge the domain gap. MAESTRO test performance remains stable (75.45 $\to$ 76.23 Frame F1), demonstrating that incorporating heterogeneous unlabeled audio does not degrade in-domain transcription.
\begin{figure*}[tb]
\centering
\includegraphics[width=\linewidth]{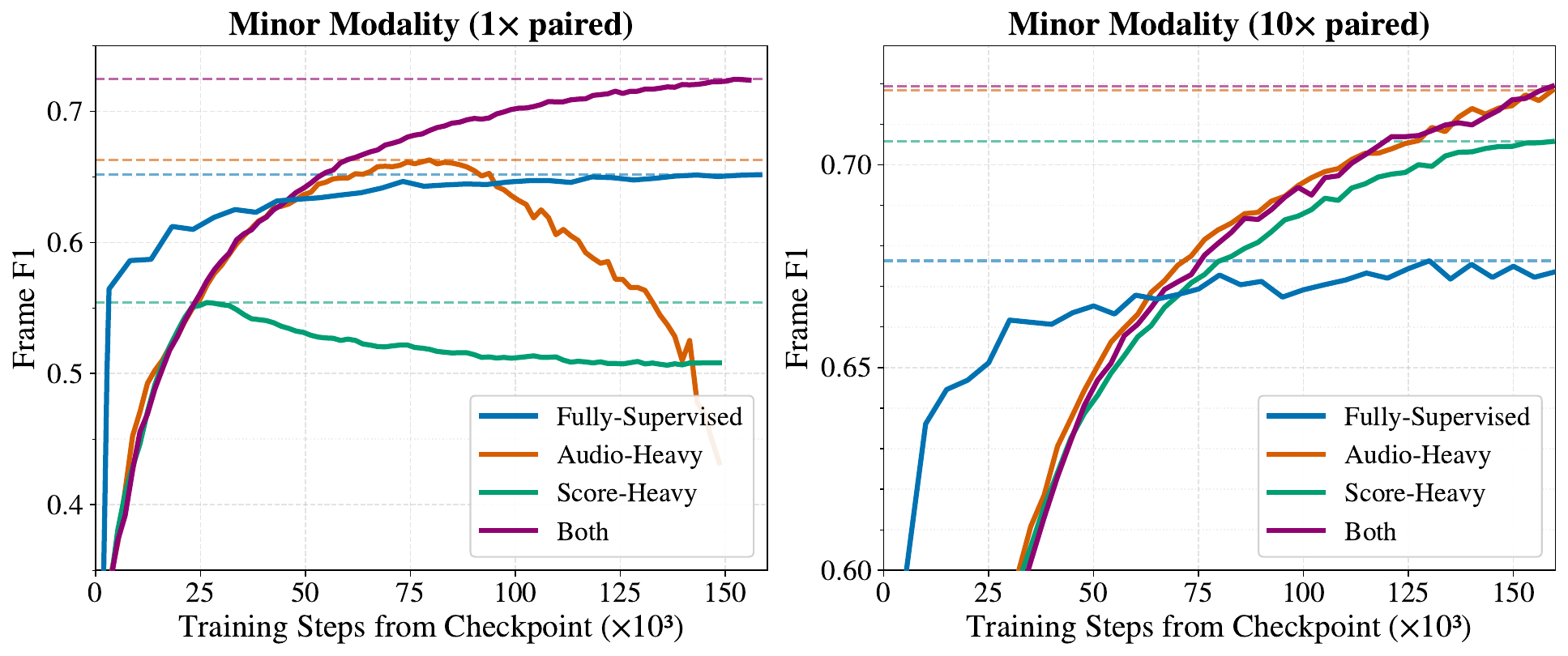}
\caption{Validation Frame F1 during training under 1.6h of paired supervision, comparing the effect of unpaired data modality and balance. \textbf{Audio-Heavy} uses 90$\times$ more unpaired audio than paired data; \textbf{Score-Heavy} uses 90$\times$ more unpaired score. The \textit{minor modality} refers to the less-dominant unpaired source (score in Audio-Heavy, audio in Score-Heavy). \textbf{Left}: minor modality matches the paired budget (1$\times$); \textbf{Right}: minor modality is scaled up to 10$\times$ the paired budget. \textbf{Both} (90$\times$ audio + 90$\times$ score) and the 
\textbf{Fully Supervised} baseline are identical across panels; visual differences reflect differing y-axis scales. Increasing the minor modality proportion stabilizes Audio-Heavy training and prevents the collapse (severe overfitting) observed in the left panel.}
\label{fig:training_curve}
\vspace{-16pt}
\end{figure*}
\vspace{-3pt}
\subsection{Effect of Unpaired Data Modality}
We investigate how the modality of unpaired data affects both generalization and the tendency to overfit. Table~\ref{tab:overfit} first demonstrates the regularization effect of unpaired data in an extreme setting: training on only 6 minutes ($\approx$0.05\% of MAESTRO) of paired data leads to severe overfitting (98.39 Train F1 vs.\ 27.82 Test F1). Adding 60 minutes of unpaired data substantially reduces this gap, demonstrating that cycle consistency constrains the model to learn representations that generalize across a broader distribution of examples, even without explicit alignment signal. Table~\ref{tab:unpaired_modality} and Figure~\ref{fig:training_curve} reveal a clear asymmetry between modalities: unpaired audio consistently outperforms unpaired score at matched budgets (72.46 vs.\ 70.51 Frame F1), suggesting that acoustic diversity provides a stronger cycle consistency signal than symbolic coverage alone.
 
However, the proportion of the minor modality critically affects training stability. 
Figure~\ref{fig:training_curve} compares two settings under 1.6h of paired supervision: the \textbf{left panel} shows a minor modality at 1$\times$ the paired budget; the \textbf{right panel} scales the minor modality up to 10$\times$. With only 1$\times$ minor modality, Audio-Heavy (1\,:\,90\,:\,1 paired\,:\,audio\,:\,score) collapses in later training stages (severe overfitting) despite strong initial performance, while Score-Heavy (1\,:\,1\,:\,90) remains stable but underperforms. Increasing the minor modality to 10$\times$ eliminates this collapse entirely: Audio-Heavy (1\,:\,90\,:\,10) and Both (90$\times$ audio + 90$\times$ score) converge stably above 72.00 Frame F1, while Score-Heavy (1\,:\,10\,:\,90) achieves a slightly weaker result (70.51 Frame F1), still surpassing the fully supervised baseline.

\subsection{Multi-Instrument Transcription}
\label{sec:multinstrument}
We evaluate our approach in a multi-instrument setting using MusicNet-EM, where only a small subset of recordings (0.47 hours, 3 songs) is provided as paired supervision across 11 instrument classes, and the remaining data is available only in unpaired form (Appendix~\ref{app:data}). This setting represents a low-resource regime in which both cross-modal alignment and multi-instrument generalization must be learned from minimal supervision.

Table~\ref{tab:musicnet} summarizes the results. With only 0.47 hours of paired data, the paired-only model achieves 35.38 Frame F1 and 5.25 Multi-Instrument Frame F1, indicating severe underfitting and poor instrument discrimination. Incorporating unpaired data substantially improves performance: adding both unpaired audio and unpaired scores (32.1h each) raises performance to 46.53 Frame F1 and 19.91 Multi-Instrument Frame F1.

\noindent\textbf{Unpaired data from mismatched sources still improves performance under minimal supervision.} With only 0.47h of paired data (3 songs), the model has almost no anchoring signal, yet incorporating unpaired data, drawn from two entirely different sources substantially improves performance (+11.15 Frame F1, +14.66 Multi-Instrument Frame F1). Crucially, the unpaired score and audio pools share no overlap with each other or with the paired set: scores come from the remaining MusicNet-EM recordings, while audio comes from the Gardner Museum collection, which differs in recording conditions and instrumentation. This mirrors the pattern observed in Section~\ref{sec:guitarset_train}, where unlabeled audio from a new instrument sufficed to bridge the domain gap, suggesting that cycle-consistent training can extract useful structure from heterogeneous unpaired data even in highly underdetermined, multi-instrument settings.

For reference, a fully supervised model trained on 32.6 hours of paired data achieves 76.84 Frame F1 and 67.84 Multi-Instrument Frame F1. While a substantial gap remains, the improvements from unpaired data under minimal supervision are significant given both the scarcity of paired data and the domain mismatch in the unpaired pool. Overall, these results indicate that the benefits of unpaired data extend to multi-instrument transcription, and that even a minimal paired anchor can enable learning of both pitch content and partial instrument structure in challenging, mismatched settings.
\begin{table*}[tb]
\centering
\caption{Transcription performance on MusicNet-EM. All paired data comes from MusicNet-EM. Unpaired data consists of 32.1h of MusicNet-EM scores disjoint from the paired set and 32.1h of audio collected from Isabella Stewart Gardner Museum. The fully supervised upper bound (gray) uses whole MusicNet-EM (32.6h) as paired data.}
\vspace{2pt}
\resizebox{0.75\linewidth}{!}{
\begin{tabular}{@{}ccc c@{}}
\toprule
\textbf{Paired data} & \textbf{Training Setup} & \textbf{Frame F1} & \textbf{Multi-inst Frame F1} \\
\midrule
0.47h {\footnotesize\color{gray}(3 songs)} & Paired-only                     & 35.38 & 5.25  \\
0.47h {\footnotesize\color{gray}(3 songs)} & Paired + unpaired               & 46.53 & 19.91 \\
\midrule
\color{gray}{32.6h {\footnotesize(308 songs)}} & \color{gray}{Paired-only}   & \color{gray}{76.84} & \color{gray}{67.84} \\
\bottomrule
\end{tabular}}
\vspace{-5pt}
\label{tab:musicnet}
\end{table*}

\section{Conclusion}
\vspace{-8pt}
We showed that the vast stores of unpaired audio and symbolic scores, freely available but previously untapped, can be turned into an effective training signal for music transcription. Cycle-consistent learning provides a principled framework for leveraging unpaired audio and symbolic data, but requires a minimal anchor: without any paired supervision, training admits pitch-shifted solutions that satisfy cycle constraints without recovering correct transcription. As little as 1.6 hours of paired data suffices to stabilize training and recover 86.3\% of fully supervised performance on MAESTRO.

We show that the modality of unpaired data matters: unpaired audio contributes more strongly than unpaired scores at matched budgets, suggesting that acoustic diversity is a primary driver of cross-modal alignment. Beyond modality, the composition of the unpaired pool affects both generalization and training stability, with a modest proportion of the complementary modality sufficient to prevent collapse.

Finally, we show that incorporating unlabeled audio from an unseen instrument substantially improves zero-shot transcription of that instrument (+10 Frame F1 on GuitarSet) without any paired target-domain supervision, and outperforms a fully supervised out-of-domain model. This improvement does not come at the cost of in-domain performance: MAESTRO transcription remains stable throughout. We observe a similar trend in multi-instrument settings, where incorporating out-of-domain unpaired audio yields consistent gains despite distribution mismatch, suggesting that these benefits extend beyond single-instrument transcription, though further study is needed to better understand under what conditions such cross-domain unpaired data is most beneficial.

\ack{We thank SoundPatrol, Inc. for a research gift and acknowledge generous support from a Cornell donor. JL is supported by a Google PhD Fellowship. This work was also supported by AI-MI, NSF Award DMR-2433348, and the NYP-Cornell Cardiovascular AI Collaboration at New York-Presbyterian. We gratefully acknowledge the research computing resources provided by the Empire AI Consortium, with support from the State of New York, the Simons Foundation, and the Secunda Family Foundation.}

\bibliographystyle{abbrvnat}
\bibliography{mybib}

\newpage
\appendix

\section{Dataset Details}
\label{app:data}

\noindent\textbf{Input representation.}
Audio is represented as a log-CQT magnitude spectrogram $x_C \in 
\mathbb{R}^{T \times B}$ computed at the native sample rate of each recording, with hop size $\lfloor \text{sr} / 50 \rfloor$ to achieve 50\,fps. Symbolic scores are represented as binary note activity matrices $x_M \in \{-1,+1\}^{T \times 88}$ at the same frame rate. Both modalities are segmented into non-overlapping 256-frame ($\approx$5.12\,s) chunks. For paired examples, scores are padded or trimmed to match the audio-derived frame count before chunking.

\noindent\textbf{MAESTRO.}
We use MAESTRO v2.0.0~\cite{hawthorne2018enabling}. Audio is converted to a log-CQT with 88 bins at 12 bins per octave, $f_{\min}$=A0, at 50\,fps (hop size 882 samples at 44.1\,kHz, 960 samples at 48\,kHz). MIDI is rendered as a binary piano roll at the same frame rate, binarized and mapped to $\{-1,+1\}$. We sample $\{0.1, 1, 5, 10\}\%$ of the training split as paired supervision (corresponding to paired-to-unpaired ratios of $\{1\!:\!1000, 1\!:\!100, 1\!:\!19, 1\!:\!9\}$) and treat the remainder as unpaired.

\noindent\textbf{GuitarSet.}
GuitarSet~\cite{xi2018guitarset} contains 360 paired guitar recordings spanning five musical styles (Jazz, Bossa Nova, Rock, Singer-Songwriter, Funk). It is used 
for two purposes: zero-shot cross-instrument evaluation, and as a source of unlabeled target-domain audio in domain adaptation experiments. We follow the 
style-progression split from MT3~\cite{gardner2021mt3}: progressions 1 and 2 for training, progression 3 for validation. Microphone (mono-mic) audio is used. Piano 
rolls are derived from per-string JAMS annotations and binarized at 50\,fps. Audio is converted to a CQT with 352 bins at 48 bins per octave, $f_{\min}$=A0.

\noindent\textbf{MusicNet-EM.}
MusicNet-EM improves the original DTW-based MusicNet annotations~\cite{thickstun2016learning} via an EM re-annotation procedure~\cite{maman2022unaligned}. We use the version distributed with YourMT3~\cite{chang2024yourmt3}, resampled to 16\,kHz, covering 11 instrument classes: piano, harpsichord, violin, viola, cello, pizzicato strings, French horn, oboe, bassoon, clarinet, and flute. Audio is converted to a CQT with 352 bins at 48 bins per octave, $f_{\min}$=A0, 50\,fps.

\textit{Split.} 10 songs constitute a fixed test set. The remaining songs are ranked by instrument diversity; the top 8 are selected as the paired pool, of which 5 are held out for validation (seed 42) and 3 serve as supervised training pairs. The remaining 305 songs form the unpaired score pool.
\textit{Statistics}: paired train: 3 songs, 330 chunks; validation: 5 songs, 467 chunks; test: 10 songs, 286 chunks; unpaired score: 305 songs, 22,604 chunks.

\noindent\textbf{Gardner Museum (Unpaired Audio).}
We use additional recordings from the Isabella Stewart Gardner Museum as unpaired audio. Since MusicNet draws in part from the Gardner Museum collection, some recordings overlap with MusicNet-EM; to ensure that unpaired audio does not contain recordings present in the paired or test sets, we apply a two-stage deduplication filter: (1) filename-based matching on composer name and catalog identifier (e.g., ``op18''), and (2) duration matching within $\pm$2 seconds against MusicNet-EM recordings. After filtering, 532 recordings ($\approx$303\,h, 213,211 chunks) are retained, processed with the same CQT configuration as MusicNet-EM. For training, we randomly sample 32.1h from this pool to match the unpaired score budget.

\section{Model Architecture}
\label{app:arch}

\noindent\textbf{Score VAE.}
We train separate Score VAEs for MAESTRO (piano-only) and MusicNet-EM (multi-instrument). Both follow the same architecture: a 2D convolutional U-Net-style encoder-decoder with residual blocks and linear attention, applied at four resolution levels with channel multipliers $(1, 2, 2, 2)$ and 2 residual blocks per level. Three spatial downsampling steps reduce the time and pitch dimensions by a factor of 8. The MAESTRO VAE takes input of shape $\mathbb{R}^{1 \times 256 \times 88}$ and produces $z_M \in \mathbb{R}^{8 \times 32 \times 11}$; the MusicNet-EM VAE takes $\mathbb{R}^{11 \times 256 \times 88}$ (instruments $\times$ frames $\times$ pitch bins) and produces the same latent shape. Reconstruction is trained with a focal loss ($\gamma{=}2.0$, $\alpha{=}0.75$) to handle class imbalance in piano rolls (predominantly silent frames). We use $\beta = 10^{-8}$, making the KL term effectively negligible and the model behave as a near-deterministic autoencoder, which preserves sharp note boundaries. VAE weights are frozen during all subsequent cycle-consistency training, using EMA-averaged weights for encoding.

\noindent\textbf{Generators.}
$G$ is a convolutional encoder that progressively maps CQT spectrograms to the score latent space through five downsampling stages with channel multipliers $\{1, 2, 2, 4, 8\}$ and base width $\mathit{ngf}=32$, incorporating residual blocks throughout. Stages 0-1 (highest resolution) use no attention, while stages 2-4 additionally apply linear attention~\cite{katharopoulos2020transformers}. The per-stage strides are $(2,2),(2,2),(2,2),(1,2),(1,2)$, reflecting the asymmetric spatial dimensions of the CQT input ($256\times352$) and the latent target ($32\times11$). $F$ is the symmetric decoder that reconstructs CQT spectrograms from latent representations using five upsampling stages with the same channel schedule and attention pattern, with a final $\tanh$ activation to produce outputs in $[-1, 1]$.

\noindent\textbf{Discriminators.}
Both discriminators follow a multi-scale design in which each scale operates on a $2\times$ average-pooled version of the previous input. The CQT discriminator $D_C$ uses three scales, each implemented as a three-layer convolutional network with anisotropic kernels of size $3\times5$, stride $(2,2)$, and $\mathit{ndf}=64$ base channels that double per layer up to 128. The score discriminator $D_S$ uses two scales, each a five-layer network with $3\times3$ kernels, stride $(1,1)$, and the same channel schedule. Both discriminators use weight-normalised residual blocks throughout and are trained with the least-squares GAN objective~\cite{mao2017least}. In addition to the adversarial loss, a feature-matching loss~\cite{wang2018high, kumar2019melgan} with weight $0.1$ is computed between the intermediate activations of $D$ on real and cycle-reconstructed samples.

\section{Training Details}
\label{app:training}
\noindent All inputs are normalized to $[-1, 1]$, and EMA weights are used for all reported metrics in the training.
\noindent All models were trained on a single NVIDIA A6000 GPU (48GB VRAM). Training the single-instrument model required approximately 4 days, while the multi-instrument model required approximately 6 days. 
\begin{table}[h]
\centering
\caption{Training hyperparameters.}
\begin{tabular}{@{}llll@{}}
\toprule
\textbf{Hyperparameter} & \textbf{Score VAE} & 
\textbf{Ours (single-inst)} & \textbf{Ours (multi-inst)} \\
\midrule
Optimizer & Adam & AdamW & AdamW \\
Learning rate & $2\times10^{-4}$ & $4\times10^{-4}$ & $2\times10^{-4}$ \\
$\beta_1, \beta_2$ & default & 0.9, 0.99 & 0.9, 0.99 \\
LR schedule & Cosine & Cosine & Cosine \\
LR warmup steps & 1,000 & 2,000 & 2,000 \\
Gradient clipping & norm 1.0 & norm 1.0 & norm 1.0 \\
Batch size & — & 64 & 32 \\
Steps / epochs & 200 ep. & 1M & 500k \\
Precision & — & bf16-mixed & bf16-mixed \\
EMA decay & 0.99 & 0.99 & 0.99 \\
\midrule
$\lambda_{\text{cyc}}$ & — & 5.0 & 5.0 \\
$\lambda_{\text{fm}}$ & — & 1.0 & 1.0 \\
$\lambda_{\text{sup}}$ & — & 1.0 & 1.0 \\
$\beta$ (KL weight) & $10^{-8}$ & — & — \\
Adv. loss warmup & — & 500 steps & 2,000 steps \\
Image pool size & — & 128 & 128 \\
Pool sampling ratio & — & 50\% & 50\% \\
\bottomrule
\end{tabular}
\label{tab:hyperparams}
\end{table}

\end{document}